# The Standard Model of Particle Physics

Tom W.B. Kibble
Blackett Laboratory, Imperial College London

**Abstract**
   This is a historical account from my personal perspective of the development over the last few decades of the standard model of particle physics.  The model is based on gauge theories, of which the first was quantum electrodynamics, describing the interactions of electrons with light.  This was later incorporated into the electroweak theory, describing electromagnetic and weak nuclear interactions.  The standard model also includes quantum chromodynamics, the theory of the strong nuclear interactions.  The final capstone of the model was the Higgs particle discovered in 2012 at CERN.  But the model is very far from being the last word; there are still many gaps in our understanding.

**Atoms and their constituents**
   Since prehistoric times, people have been asking:  What are we made of?  What are the basic constituents of the world around us?  And how do they interact?  The atomic theory, that matter is composed of indivisible atoms, goes back at least to the fifth century BC, when it was championed by Leucippus and Democritus.  In the nineteenth century, chemists, beginning with John Dalton, gradually established the reality of atoms, still seen as indivisible.
   But we now know they are far from indivisible.  Most of the mass of an atom is concentrated in the tiny nucleus surrounded by a cloud of electrons.  The nucleus in turn is formed of protons and neutrons, but these are not elementary either.  They are actually composed of even smaller particles, called quarks.  So far as we now know that is as far as it goes: electrons and quarks really are elementary.  But who knows how the theory will develop in the future?
   Gradually over the last century we have built up what we now call the standard model of particle physics, which describes the particles we now believe to be fundamental and their interactions.  The theory has proven amazingly successful, accurately describing almost all of the empirical data, though there are a few tiny niggling discrepancies.  In this article, I am aiming to give a historical account, from my personal perspective, of how this model developed.  It is a story of encounters with many intractable obstacles and ingenious ways of avoiding them.
   Describing this model is not easy, partly because it is really quite complex, but even more because we lack the language.  One thing we have learned is that small particles are not at all like tiny versions of large ones.  They behave in ways that seem in the light of ordinary experience truly bizarre.  So analogies with things we are familiar with tend to be flawed and misleading.  Nevertheless, I shall try.
   The first bit of weirdness concerns waves and particles, two very different things in our familiar world.  A debate raged for 200 years: is light a wave or is it formed of particles?  The answer of course is both.  As Thomas Young showed, it is certainly a wave; if a light beam strikes two parallel slits then on a screen behind them it creates strips of light and dark where the waves reinforce each other or cancel out.  But Einstein's work on the photoelectric effect (for which he got the Nobel Prize) demonstrated that when light is absorbed energy can only be extracted in discrete quanta, called photons, with energy $E$ related to the frequency $f$ of the light by $E = hf$, where $h$ is Planck's constant. So whether light behaves as a wave or a particle depends on what we choose to look at.

The same, we now know, is true of electrons and other particles; they exhibit wavelike properties. They too produce interference fringes when passed through two parallel slits. That is weird: if they are particles, then, one would imagine, they must pass through one slit or the other. Yet if we arrange a device to observe which slit they go through the pattern disappears. But though weird, the wave properties of electrons are now well understood and indeed practically useful. The wavelengths involved are tiny, so if we want much finer resolution than an ordinary optical microscope provides, we use an electron microscope. And indeed to see detail on a sub-atomic scale we need even more energetic probes, which is why we need big particle accelerators.

Curiously, Planck's constant $h$ turns up in another apparently unrelated setting. Most of the known particles have an intrinsic spin, corresponding to an angular momentum which is always an integral or half-integral multiple of $h/2\pi$, a quantity for which we use the special symbol $\hbar$. There are two very different families of particles: bosons, whose spin is $0, \hbar, 2\hbar, \cdots$ and fermions, with spin $\frac{1}{2}\hbar, \frac{3}{2}\hbar, \cdots$ .

For simplicity we talk of "spin 0", "spin ½", "spin 1" and so on. Bosons are sociable and like to occupy the same state, whereas fermions (including protons, neutrons and electrons) are forbidden from doing so by the Pauli exclusion principle, the essential basis of atomic structure.

**Quantum electrodynamics**

The standard model describes how these curious entities interact. The first piece of it to reach a mature form was quantum electrodynamics (QED) which deals with the interactions of light, or photons, with electrons. It is a field theory; each kind of particle is described by a field extending through space. Photons of course are related to the electric and magnetic fields, and there is also an electron field. Particles can be thought of as little wave packets, short trains of waves in these fields, though the analogy is far from perfect.

QED was invented in the 1930s and gave good results, but also ran into a serious problem. The standard method of calculation, called perturbation theory, involves successive approximations. The first approximation gave results usually within a few percent of the correct answer. The problem was that the higher approximations, intended to give more accurate answers, actually gave the answer infinity! The solution to that problem emerged just after the second world war. It was found independently by three people, in 1947 by Richard Feynman and by Julian Schwinger in the USA and in fact earlier, in 1943, by Sin-Itiro Tomonaga working all alone in wartime Japan. It was called renormalization theory. Essentially it amounted to gathering all the infinities together into corrections to the mass and electric charge of the electron. This juggling with infinite quantities still seems a rather suspect procedure, but it works; after renormalization the results obtained were finite and agreed to astonishing accuracy with the experiments. QED rapidly became the most accurately verified theory in the history of physics.

So what next? By the time I had the good fortune to join Abdus Salam's group at Imperial College in 1959, QED was a well-established theory. But there are other forces besides the electromagnetic, and physicists started searching for equally successful models of those.

We distinguish four kinds of forces. Two of these, the electromagnetic and gravitational, are familiar because they are long-range forces, diminishing as the inverse square of the distance. In QED the electromagnetic force between two electrons can be thought of as due to the exchange of photons between them. It is long-range because the photon is massless, by which we mean it has no rest-mass. A massive particle with rest-mass $m$ has a minimum energy, when it is at rest, given by Einstein's famous relation $E = mc^2$. But a photon is never at rest and its energy can be as small as one likes.

Gravitational forces are long-range too, supposedly transmitted by another massless particle, the graviton. But on the scale of particle physics, gravitational forces are tiny; the gravitational attraction between a pair of protons is weaker than the electrostatic repulsion by a factor of a trillion trillion trillion. Gravitons are therefore at the moment (and perhaps for ever) completely undetectable, though physicists hope to be able to detect gravitational waves within the next few years.

**The search for symmetries**

A principal goal of theoretical particle physicists after the success of QED was to understand the two other kinds of forces which are less familiar because they are short-range, negligible on scales larger than an atomic nucleus. These are the strong and weak nuclear forces. The strong force is responsible for binding protons and neutrons together in atomic nuclei. The weak force appears in one form of radioactivity, nuclear beta decay, in which a neutron changes into a proton, an electron and a very elusive massless neutrino. It also plays a vital role in the energy generating mechanism in the Sun.

Experimental particle physics also saw rapid development. Before the second world war we knew of a handful of presumably elementary particles, but when after the war physicists went back to studies of cosmic rays and later of collisions in particle accelerators, they rapidly discovered a huge zoo of particles, a hundred or more, so people started to ask, could all these really be elementary? To make sense of this zoo, physicists looked for patterns, rather as chemists had done a century earlier in developing the periodic table. This revealed a lot of similarities and approximate symmetries.

Before the war Werner Heisenberg had suggested a symmetry between protons and neutrons. He pointed out that they are very similar, with the same strong interactions, the same spin (½) and almost the same mass, and suggested that they might be regarded as two different states of a single entity, called the nucleon. He also proposed a symmetry in which one can as it were rotate one into the other without changing anything. This symmetry is now called isospin, not because it has anything to do with real spin, but because of a mathematical analogy with the symmetry between states of a spinning electron. Of course, this is not an exact symmetry. Protons and neutrons do differ, because the proton has an electric charge and the neutron does not; the symmetry is broken by electromagnetism. But it is nevertheless a good approximation, for example in classifying energy levels of light atomic nuclei, because at short distances the strong interactions are much stronger than the electromagnetic.

Later, in 1961, it was found that many of the newly discovered particles could be arranged in two-dimensional patterns suggesting a wider but even more approximate symmetry. This was discovered independently by Murray Gell-Mann and by Yuval Ne'eman, at the time a student of Salam's at Imperial College. Gell-Mann called the symmetry the "eight-fold way", because of its characteristic patterns of octets.

**Gauge theories**

Where were we to look for theories of the strong and weak interactions? Some people argued for abandoning field theory altogether and advocated an alternative, called S-matrix theory. But Salam and others of like mind felt that the answer lay in a particular kind of field theory, a gauge theory. QED is the simplest example of a gauge theory. That means it has a particular type of symmetry, whose simplest manifestation is that it is differences in electric voltages that matter, not absolute voltages.

The first gauge theory beyond QED was proposed in 1954 by Chen-Ning Yang and Robert Mills, and incorporated Heisenberg's isospin symmetry. The same theory was actually written down independently by a student of Salam's, Ronald Shaw, though he never published it except as a Cambridge University PhD thesis. It was intended as a theory of strong interactions. In the end it turned out not to be the correct theory, but it is nevertheless the basis for all later work on gauge theories.

There were also proposals for a gauge theory of weak interactions, the first by Schwinger in 1956. It had been discovered that just as electromagnetic processes proceed via exchange of photons, so the weak interactions could be seen as mediated by particles called $W^+$ and $W^-$, where the superscripts indicate positive and negative electric charge. In many ways these particles were similar to the photon — in particular they all have spin 1 — but with one major difference. To explain the extremely short range of the weak interactions it had to be assumed that the $W$ particles are very massive, in fact about 100 times heavier than a proton. Schwinger suggested they might also be the "gauge bosons" of a gauge theory, and indeed that there might be a unified theory of weak and electromagnetic interactions with the $W^+$, $W^-$ and photon (conventionally denoted by $\gamma$) appearing symmetrically. But of course this symmetry could not be exact; because of the big difference in mass, it had to be severely broken.

Other proposals followed. In 1961, Sheldon Glashow proposed a modified theory. To solve another problem with Schwinger's scheme, concerning mirror symmetry, he added a fourth gauge boson, $Z^0$, electrically neutral like the photon. Salam and his collaborator, John Ward, proposed essentially the same theory in 1964, apparently unaware of Glashow's work.

**The problem of mass**

But there was a big problem with all these proposals. Some mechanism had to be found to break the symmetry, leaving the photon massless while giving large masses to the other gauge bosons. Just putting in masses by hand spoiled the nice properties of gauge theories, and indeed rendered them unrenormalizable; the process of renormalization could not be used to make them give finite answers. So people began to ask, could the symmetry be broken spontaneously? Spontaneous symmetry breaking is a well-known and ubiquitous phenomenon. It means that the underlying theory remains symmetric, but the particular realization we are dealing with is not. It frequently occurs at a phase transition, as from liquid to solid, where we move from a phase where the symmetry is manifest to one where it is hidden. If you place a circular bowl of water on a table, it looks exactly the same from every direction; it has rotational symmetry. But when it freezes, the ice crystals will line up in particular directions, breaking the symmetry. The breaking is spontaneous in the sense that we cannot predict in advance which direction will be chosen, unless we know of external features that already break the symmetry, such as imperfections in the bowl.

However, this did not immediately solve the problem, because it was widely believed that in any theory compatible with Einstein's special theory of relativity spontaneous symmetry breaking would always lead to the appearance of unwanted massless "Nambu-Goldstone bosons", unwanted because no one had seen any such particles, though they should have been easy to detect. These particles correspond to waves in the direction of the symmetry breaking. When Steven Weinberg visited Imperial College on sabbatical in 1961, he and Salam spent a lot of time discussing this problem. Their unhappy conclusion, that such particles are inevitable, was the content of the "Goldstone theorem", published together with Jeffrey Goldstone.

The escape from this obstacle was found in 1964 by three groups of people independently, firstly François Englert and Robert Brout from Brussels, then Peter Higgs from Edinburgh, and finally a couple of months later by two American colleagues, Gerald Guralnik and Carl

Richard Hagen, and myself at Imperial College.  The three groups all published papers in Physical Review Letters in the summer and autumn of 1964.  They approached the problem from very different perspectives, but came to essentially the same conclusion.  There is nothing wrong with the proof of the Goldstone theorem, but it relies on a very natural assumption that is nevertheless false for gauge theories.  The mechanism has been described as one in which the gauge bosons "eat" the Nambu-Goldstone bosons and hence acquire mass.  The spontaneous symmetry breaking is achieved through the action of a new scalar field, whose corresponding particles are the Higgs bosons.  "Scalar" here means that the Higgs bosons, uniquely among known elementary particles, have spin zero.

These three papers attracted almost no interest at the time.  Although by then we had in place both Glashow's unified model and the mass-generating mechanism, it took three more years for anyone to put the two together.  I wrote another paper on the subject in 1967, exploring the way the mechanism would apply to more realistic models, not just the simplest gauge theory discussed in the 1964 papers.  That work helped, I believe, to revive interest in the problem, particularly Salam's interest.  Finally a unified theory of weak and electromagnetic interactions, combining Glashow's model with the mass-generating mechanism, was proposed by Weinberg in late 1967.  Salam developed essentially the same theory independently, and presented it in lectures he gave at Imperial College in the autumn of that year, but he did not publish it until the following year.  He called it the "electroweak theory".  Both he and Weinberg believed that the theory was indeed renormalizable and therefore self-consistent, but this was only proved in 1971 in a remarkable *tour de force* by a young student, Gerard 't Hooft.  The correctness of the model was confirmed over the next twenty years by experiments at CERN and elsewhere, including the discovery of the $W^+$, $W^-$, and $Z^0$ particles in 1983.

**Quarks and gluons**

While all this was going on, there were also remarkable, though very different, developments in understanding the strong interactions.  This started with an attempt to understand the eightfold-way symmetry mentioned earlier.  It was pointed out in 1963 by Gell-Mann and independently by George Zweig that this pattern could be understood if all the strongly interacting particles were composites of three basic entities.  Picking a name from Finnegan's Wake, Gell-Mann called them "quarks", distinguished as "up", "down" and "strange" (*u*, *d*, and *s*).  The proton and neutron for example were each formed from three quarks, the proton from (*uud*) and the neutron from (*udd*).  Strange quarks, which were significantly heavier, were needed to make the more exotic members of the particle zoo.  Viewed thus, isospin was a symmetry between *u* and *d* quarks, and the eightfold way a more approximate symmetry between *u*, *d* and *s*.  One of the very strange properties of quarks is that, unlike all previously known particles, they have electric charges that are fractional multiples of the proton charge $e$, $+\frac{2}{3}e$ for *u* and $-\frac{1}{3}e$ for *d* and *s*.  How that is compatible with the fact that no one has ever seen a particle with fractional charge is a question we shall return to below.

However, there was a flaw in this neat scheme, most readily illustrated by the existence of a particle called $\Delta^{++}$, formed of (*uuu*).  Since it has spin $\frac{3}{2}$, the spins of the three *u* quarks (each ½) must be pointing in the same direction.  This means they are occupying the same state, in contradiction to the Pauli principle.  A solution to that problem was devised in 1964 by Moo-Young Han and Yoichiro Nambu and independently by Oscar Wallace Greenberg: each of the quarks comes in three different "colours", conventionally red, green and blue, although of course they have nothing to do with real colours.  All the known particles are colour neutral, containing say one quark of each colour (or else a quark and its antiparticle).

Thus the three quarks in (*uuu*) are of three different colours and so don't violate the Pauli principle. There is perfect symmetry between the different colours.

We now know that there are in fact six types of quark, not just three. The three extra are called "charm" (*c*), "bottom" (*b*) and "top" (*t*); more glamorous names for the last two, "beauty" and "truth", have not caught on. The six quarks can be arranged as three pairs, (*u,d*), (*c,s*), (*t,b*), each with the same pair of charges ($+\frac{2}{3}e, -\frac{1}{3}e$). Each of the three pairs looks essentially the same, except for the fact that they get increasingly heavy; the top quark is actually heavier than the *W* and *Z* gauge bosons. The electron too has two very similar but heavier companions, the muon ($\mu$) and the tau ($\tau$), and each of these three is paired with a massless neutrino with charge zero. So for both the quarks and what we call the leptons there is an odd three-generation structure.

Physicists took some time to be convinced that this apparently artificial solution is in fact correct. But colour proved to be the essential key to understanding the strong interactions. Han and Nambu suggested that strong interactions might be mediated by an octet of gauge bosons, called "gluons", coupled specifically to colours. This idea forms the basis of the now accepted gauge theory of strong interactions, quantum chromodynamics (QCD). This colour force induced by gluon exchange has some really odd features. Familiar forces like electromagnetism always fall off with increasing distance. But in 1973 David Gross and Frank Wilczek, and also David Politzer, showed that by contrast the colour force exhibits what they called "asymptotic freedom", namely it becomes very weak at short distances. ("Asymptotic" here indicates that the effect is seen in collisions at very high energy.) This is very fortunate for theorists, because it means that the usual perturbation-theory technique can be used to calculate these high-energy effects. Why no one has seen any free quarks or gluons is explained by the obverse of this phenomenon, called "confinement", namely that the colour force between particles that are not colour-neutral grows at large distance; it is as if the quarks inside a proton or neutron are bound together by elastic threads and can never escape.

**The Higgs and beyond**

This may seem a very strange theory but it is now well established; QCD and the electroweak theory together constitute the standard model, with spin-½ leptons and quarks and spin-1 gauge bosons. It has been tested by innumerable experiments over the last forty years and been thoroughly vindicated.

Until recently there was however a gap, the Higgs boson. Back in 1964, the existence of this extra particle was seen as a relatively minor feature; the important thing was the mechanism for giving masses to gauge bosons. But twenty years later, it began to assume a special significance as the only remaining piece of the standard-model jigsaw that had not been found. Finding it was one of the principal goals of the large hadron collider (LHC) at CERN. This is the largest piece of scientific apparatus every constructed, a precision instrument built in a huge 27 km-long tunnel straddling the French-Swiss border near Geneva — a truly remarkable piece of engineering. Protons are sent round in both directions, accelerated close to the speed of light, and allowed to collide at four crossing points around the ring. At two of these are large detectors, Atlas and CMS, also marvels of engineering, that over a period of twenty years have been designed, built and operated by huge international teams of physicists and engineers. In 2012 this mammoth effort paid off, with the unequivocal discovery by both teams of the Higgs boson.

So is this the end of the story? Surely not. The standard model can hardly be the last word. It is marvellously successful, but far from simple. It has something like 20 arbitrary parameters, things like ratios of masses and coupling strengths, that we cannot predict and that seem to have no obvious pattern to them. Moreover there are many features for which we have no explanation. Why for both quarks and leptons are there three generations with very similar properties but wildly varying masses? Why do quarks come in three colours? One theory is that all these choices are random. There may have been many big bangs, each producing a universe with its own set of parameters. Most of those universes would probably be devoid of life. But that is for many a profoundly unsatisfactory answer; we certainly hoped for a more predictive theory!

On the observational side, there are still many things we cannot explain. What is the nature of the dark matter in the universe? Why does the universe contain more matter than antimatter — leptons and quarks rather than antileptons and antiquarks? Moreover there are a few points on which the standard model definitely does not agree with observation. In particular, in the standard model the neutrinos are strictly massless. But we now know that do in fact have non-zero, albeit very tiny, masses. We really have no idea why.

Finally, there is the elephant in the room: gravity, which does not appear at all in the standard model. It is in fact very difficult to reconcile our best theory of gravity, Einstein's general theory of relativity, with quantum theory. That is a problem we have been struggling with for the best part of a century. There are hopes that string theory, or its more modern realization, M-theory, may successfully unite the two, but that effort has been going on for decades without as yet reaching a conclusion. At any rate it does appear that there is a lot more for theoretical physicists to do!

**Further Reading**
Sample, I. (2010) *Massive: the Hunt for the God Particle* (London: Virgin Books)
Close, F. (2011) *The Infinity Puzzle* (Oxford: Oxford University Press)
Carroll, S. (2013) *The Particle at the End of the Universe* (London: One World Publications)

**Acknowledgement**
This article is the text of an invited talk given at the 25th Anniversary Meeting of the Academia Europaea held in Wrocław in September 2013. I am grateful to the organizers for the invitation.